\documentclass[11pt,a4paper]{article}
\usepackage{graphicx,epsfig}
\usepackage{subeqn} 

\newcommand{\rmd}{{\rm d}}

\newcommand{\bk}{{\bf k}}

\newcommand{\cH}{{\mathcal H}}

\begin{document}

\begin{center}
{\Large PROPERTIES OF FRACTIONAL EXCLUSION STATISTICS IN INTERACTING PARTICLE SYSTEMS}

\bigskip \bigskip

\small
DRAGO\c S-VICTOR ANGHEL

\bigskip

\footnotesize
{\em Department of Theoretical Physics, Horia Hulubei National Institute for Physics and Nuclear Engineering (IFIN-HH), 407 Atomistilor, Magurele-Bucharest 077125, Romania\\ E-mail: dragos@theory.nipne.ro}

\bigskip

\small

(Received \today ) 
\end{center}

\bigskip

\footnotesize

{\em Abstract.\/} 
We show that fractional exclusion statistics is manifested in 
general in interacting systems and 
we discuss the conjecture recently introduced (J. Phys. A: Math. Theor. 
{\bf 40}, F1013, 2007), according to which if in a thermodynamic system 
the mutual 
exclusion statistics parameters are not zero, then they have to be 
proportional to the dimension of the Hilbert space on which they act. 
By using simple, intuitive arguments, but also concrete calculations 
in interacting systems models, we show that this conjecture is not 
some abstract consequence of unphysical modelling, but is 
a natural--and for a long time overlooked--property of fractional 
exclusion statistics. 
We show also that the fractional exclusion statistics is the 
consequence of interaction between the particles of the system and it is due 
to the change from the description of the system in terms of 
free-particle energies, to the description in terms of the quasi-particle 
energies. From this result, the thermodynamic equivalence 
of systems of the same, constant density of states, but any exclusion 
statistics follows immediately. 

\bigskip

{\em Key words:\/} Quantum statistics, ensemble equivalence, thermodynamic equivalence

\normalsize

\section{INTRODUCTION \label{intro}}

Fractional exclusion statistics (FES) is a generalization of Bose 
and Fermi statistics, introduced by Haldane in Ref. 
\cite{PhysRevLett.67.937.1991.Haldane}, and with its thermodynamic 
properties calculated mainly by Isakov \cite{PhysRevLett.73.2150.1994.Isakov} 
and Wu \cite{PhysRevLett.73.922.1994.Wu}. 
The concept have been applied to a large number of systems of interacting 
particles \cite{PhysRevLett.81.489.1998.Carmelo,PhysRevLett.72.600.1994.Veigy,JPhysB33.3895.2000.Bhaduri,PhysRevB.60.6517.1999.Murthy,NuclPhysB470.291.1996.Hansson,IntJModPhysA12.1895.1997.Isakov,PhysRevLett.73.3331.1994.Murthy,PhysRevLett.74.3912.1995.Sen,PhysRevLett.86.2930.2001.Hansson,NuclPhysB.572.547.2000.Bouwknegt,arXiv:0712.2174v1.Ouvry}
and describes quasiparticles 
that exist in finite dimensional Hilbert spaces--in general, quasiparticles 
in a finite region of condensed matter; the Hilbert spaces are extensive, 
increasing proportionally to the size of the condensed matter region 
\cite{PhysRevLett.67.937.1991.Haldane}. Quasiparticles that exist in 
these Hilbert spaces are called in general \textit{species}--each subspace 
contains one species. 
Let us denote the Hilbert spaces by $\cH_i$, with $i=0,1,\ldots$, 
each of them containing $N_i$ particles and $G_i$ available states 
\cite{JPhysA.40.F1013.2007.Anghel}. Then, 
by increasing the  number of particles of one species, say $N_i$ increases 
to $N_i+\delta N_i$, the number of available states in any of the Hilbert 
spaces changes by $\delta G_j=-\alpha_{ji}\delta N_i$. 
The proportionality factors, $\alpha_{ij}$, are called \textit{direct} 
(when $i=j$) and \textit{mutual} (when $i\ne j$) \textit{exclusion statistics 
parameters}. 

The simplest examples are the Bose and Fermi statistics. For ideal bosons, 
$\alpha_{ij}=0$ for any $i$ and $j$, whereas for ideal fermions 
$\alpha_{ii}=1$ for any $i$ and $\alpha_{ij}=0$ for any $i$ and $j$, 
if $i\ne i$. 

\section{PROPERTIES OF THE MUTUAL EXCLUSION STATISTICS PARAMETERS}

\begin{figure}[t]
\begin{center}
\resizebox{50mm}{!}{\includegraphics{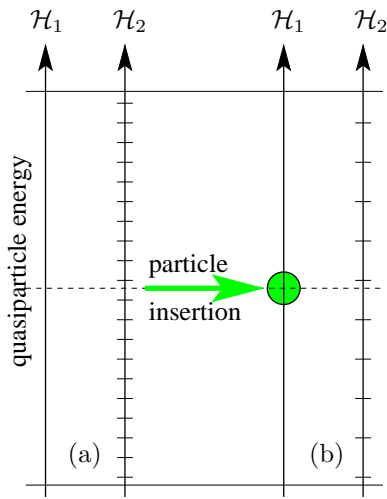}}
\end{center}
\caption{In (a), $\cH_1$ and $\cH_2$ are two Hilbert spaces defined 
by the ranges of quasiparticle energies (the energy levels in $\cH_1$ 
are not relevant, and therefore are not represented). 
In (b), extra particles are inserted into $\cH_1$, which changes 
the density of states in $\cH_2$ and 
therefore the number of allowed single-particle states. 
This is a manifestation of fractional exclusion statistics.}
\label{mutual}
\end{figure}

The FES (other than Bose and Fermi statistics) is the result of interaction 
between quasiparticles. 
The finite dimensional Hilbert spaces involved, $\cH_i$, 
are formed of quasiparticle wavefunctions corresponding to eigenvalues 
contained in finite regions of the phase space.  For example Isakov 
\cite{PhysRevLett.73.2150.1994.Isakov} and Iguchi 
and Sutherland \cite{PhysRevLett.80.1698.1998.Iguchi,PhysRevLett.85.2781.2000.Iguchi} define the Hilbert spaces by the wavevector eigenvalues, $\bk$, whereas 
Murthy and Shankar \cite{PhysRevLett.73.3331.1994.Murthy}, Sen and 
Bhaduri \cite{PhysRevLett.74.3912.1995.Sen}, Hansson, Leinaas, and 
Viefers \cite{PhysRevLett.86.2930.2001.Hansson} define the Hilbert spaces 
by the quasiparticle energy eigenvalues, $\tilde\epsilon$. In all the cases, 
the eigenvalues of the wavefunctions contained in any of the Hilbert spaces 
belong to a finite range (or a multi-dimensional volume); 
therefore the quasiparticle-quasiparticle interaction, which 
changes the density of states (DOS) in the phase-space, changes also the number 
of quasiparticle states in the range, leading to FES 
\cite{submitted.FESinteraction} (see figure \ref{mutual}). 

\begin{figure}[t]
\begin{center}
\includegraphics{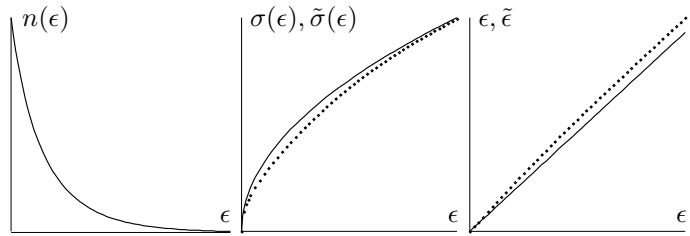}
\end{center}
\caption{The DOS on the quasiparticle energy 
axis, $\tilde\sigma[\tilde(\epsilon)]$--the point line in the middle plot--, 
and the quasiparticle energy, $\tilde\epsilon$--the point line in the 
right plot--for a bosonic single particle energy levels population, 
$n(\epsilon)=[\exp(\epsilon+1)-1]^{-1}$, a DOS 
$\sigma(\epsilon)=\sqrt{\epsilon}$ and a constant interaction potential, 
$V(\epsilon,\epsilon')=1$. In the right plot we show both, 
$\tilde\epsilon(\epsilon)$ (point line) and $\epsilon(\epsilon)$ (solid line) 
for comparison.}
\label{MSmodel}
\end{figure}

Let's take for example a system where the quasiparticle energies 
may be written as \cite{PhysRevLett.73.3331.1994.Murthy,PhysRevLett.74.3912.1995.Sen,PhysRevLett.86.2930.2001.Hansson,PhysRevB.60.6517.1999.Murthy,JPA35.7255.2002.Anghel,RomRepPhys59.235.2007.Anghel,submitted.FESinteraction} 
\begin{equation}
\tilde\epsilon_i = \epsilon_i + \sum_{j=0}^{i-1} V_{ij}n_j +\frac{1}{2}V_{ii}n_i,
\label{epstilgen}
\end{equation}
where $\epsilon_i$ are the energies of the noninteracting particles;
The total energy of the system is $E = \sum_{i=0}^\infty n_i\tilde\epsilon_i$. 
We assume that the system in large enough--and the energy levels 
dense enough--to introduce the 
(quasi)continuous density of states, $\sigma(\epsilon)$ and write 
equation (\ref{epstilgen}) as 
\begin{equation}
\tilde\epsilon = \epsilon + \int_{0}^\epsilon V(\epsilon,\epsilon')
n(\epsilon')\sigma(\epsilon')\rmd\epsilon', 
\label{epstilgenint}
\end{equation}
where we replaced the subscripts $i$ and $j$ of $V_{ij}$ by the corresponding 
energy levels, $\epsilon$ and $\epsilon'$, respectively. In the 
variable $\tilde\epsilon$, the density of states is different from 
$\sigma(\epsilon)$ and we shall denote it by 
$\tilde\sigma(\tilde\epsilon)$. To calculate $\tilde\sigma(\tilde\epsilon)$, 
we take a small interval, $\delta\epsilon$, containing 
$\sigma(\epsilon)\cdot\delta\epsilon$ states, and transform it into the 
interval $\delta\tilde\epsilon$, which, obviously, will contain the same 
number of states. Dividing the number of states by the energy interval 
and using equation (\ref{epstilgenint}), we find 
\begin{equation}
\tilde\sigma[\tilde\epsilon(\epsilon)] = \frac{\sigma(\epsilon)}
{1 + \int_{0}^{\epsilon}\frac{\partial V(\epsilon,\epsilon')}{\partial\epsilon}
\sigma(\epsilon')n(\epsilon') d\epsilon' 
+ V(\epsilon,\epsilon'\nearrow\epsilon)\sigma(\epsilon) n(\epsilon)} 
\label{sigma_disc}
\end{equation}
In figure \ref{MSmodel} we show an example. In some arbitrary, dimensionless 
units, with the population of the single particle energy levels shown in 
the left plot, the DOS $\sigma(\epsilon)$ shown as the solid line in 
the middle plot, and constant interaction potential, 
$V(\epsilon,\epsilon')\equiv V$, we calculate 
$\tilde\sigma[\tilde\epsilon(\epsilon)]$ (point line in the middle plot) and 
$\tilde\epsilon$ (point line in the right plot)--for the concrete calculations 
of figure \ref{MSmodel} we choose $V=1$, $\sigma(\epsilon)=\sqrt{\epsilon}$, 
and $n(\epsilon)=[\exp(\epsilon+1)-1]^{-1}$. 

\begin{figure}[t]
\begin{center}
\includegraphics{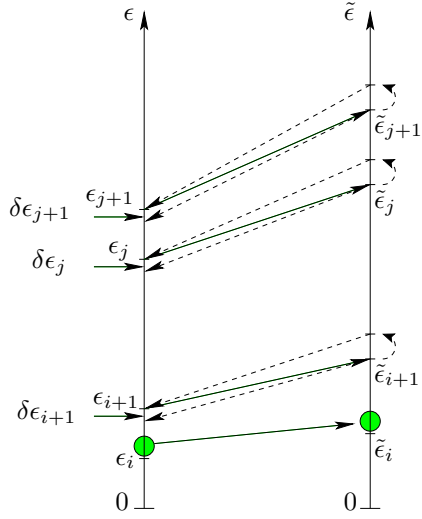}
\end{center}
\caption{The intervals $[\epsilon_0,\epsilon_1], \ldots,
[\epsilon_{j-1},\epsilon_j], \ldots$, on the $\epsilon$ axis, and the 
corresponding intervals, $[\tilde\epsilon_0,\tilde\epsilon_1], \ldots,
[\tilde\epsilon_{j-1},\tilde\epsilon_j],\ldots$ on the $\tilde\epsilon$ axis. 
The insertion of particles in the interval 
$[\tilde\epsilon_{i},\tilde\epsilon_{i+1}]$ changes the intervals 
on the $\epsilon$ axis, from $[\epsilon_{i},\epsilon_{i+1}]$, above, if 
the intervals on the $\tilde\epsilon$ axis are held fixed.}
\label{FESfig}
\end{figure}

To show how FES is manifested in the system, 
we split the quasiparticle energy axis into small intervals, 
$[\tilde\epsilon_0,\tilde\epsilon_1], \ldots,
[\tilde\epsilon_{i-1},\tilde\epsilon_i], \ldots$ (as shown in figure 
\ref{FESfig}), so that each interval 
contains large enough numbers of particles and available single particle 
states; we denote by $N(\tilde\epsilon_{i},\tilde\epsilon_{i+1})\equiv N_i$ 
the number of particles in the interval 
$[\tilde\epsilon_{i},\tilde\epsilon_{i+1}]$, 
\[
N_i\equiv N(\tilde\epsilon_{i},\tilde\epsilon_{i+1})=
\int_{\tilde\epsilon_{i}}^{\tilde\epsilon_{i+1}}
\tilde\sigma(\tilde\epsilon)n(\tilde\epsilon)\,d\tilde\epsilon
= \int_{\epsilon_{i}}^{\epsilon_{i+1}}
\sigma(\epsilon')n(\epsilon')\,d\epsilon',
\]
and, assuming that the energy intervals are small enough so that 
we can replace in the integral of equation (\ref{epstilgenint}) 
$V(\epsilon_M,\epsilon)$ by $V(\epsilon_M,\epsilon_{i-1})$ 
for any $\epsilon\in(\epsilon_i,\epsilon_{i-1}]$, we write 
\begin{equation}
\tilde\epsilon_j = \epsilon_j + \sum_{i=0}^{j-1} V(\epsilon_j,
\epsilon_{i})N(\tilde\epsilon_{i},\tilde\epsilon_{i+1}) .
\label{tildeeps_sum_interv}
\end{equation}
%
%
If we insert $\delta N_i$ particles into the interval 
$[\tilde\epsilon_{i},\tilde\epsilon_{i+1}]$ and we keep all the 
free particle energy levels $\epsilon_{j(>i)}$ unchanged, then we change 
the quasiparticle energy levels $\tilde\epsilon_{j(>i)}$ by 
$\delta\tilde\epsilon_{j(>i)}=V(\epsilon_j,\epsilon_i)\delta N_i$ 
(see the small, round, arrows on the $\tilde\epsilon$ axis in figure 
\ref{FESfig}). In this way, in all the intervals 
$[\tilde\epsilon_{j},\tilde\epsilon_{j+1}]$, with 
$j\ne i$, the $N_j$ and $G_j$ remain unchanged. 
Therefore, if we maximize the partition function with respect to the 
population of the energy intervals $[\epsilon_{i},\epsilon_{i+1}]$, 
$i=0,1,\ldots$, we have to take into account the change of 
the quasiparticle energies due to the change of the populations. 
We shall come back to this method later. 

Another method to calculate the partition function and its maximum, is to 
fix the intervals $[\tilde\epsilon_0,\tilde\epsilon_1], \ldots,
[\tilde\epsilon_{j-1},\tilde\epsilon_j], \ldots$ along the 
$\tilde\epsilon$ axis. In this case, the insertion of the $\delta N_i$ 
particles into the interval  $[\tilde\epsilon_{i},\tilde\epsilon_{i+1}]$
changes the values of $\epsilon_j$ (for $j>i$) given by equation 
(\ref{tildeeps_sum_interv}) into $\epsilon'_j$, which should be calculated 
like in \cite{submitted.FESinteraction}: 
\begin{equation}
\tilde\epsilon_j = \epsilon'_j + V(\epsilon'_j,\epsilon_i)I_i 
+ \sum_{k=0}^{i} V[\epsilon'_j,\epsilon_{k}]
N(\tilde\epsilon_k,\tilde\epsilon_{k+1}) 
+ \sum_{k=i+1}^{j} V[\epsilon'_j,\epsilon'_{k}]
N(\tilde\epsilon_k,\tilde\epsilon_{k+1}) .
\label{tildeeps_sum_interv_ip}
\end{equation}
Because $\sigma(\epsilon)$ is independent of the population, the change 
of $\epsilon_{j(>i)}$ (see the $\delta\epsilon_{j}$s on the left 
axis in figure \ref{FESfig}) leads to a change in the number of 
available single particle states in the interval 
$[\epsilon'_{j},\epsilon'_{j+1}]$ and therefore a change of the number 
of states in the interval $[\tilde\epsilon_{j},\tilde\epsilon_{j+1}]$ and 
a change of $\tilde\sigma(\tilde\epsilon)$, according to equation 
(\ref{sigma_disc}). \textit{This is the manifestation of FES}. 

The exclusion statistics parameters of the FES gas have been calculated 
in Ref. \cite{submitted.FESinteraction}. There we showed in the general 
case that these parameters obey the properties conjectured in 
Ref. \cite{JPhysA.40.F1013.2007.Anghel}, namely that the mutual exclusion 
statistics parameters are proportional to the dimension of the Hilbert 
space on which they act. Let us take as example a system of \textit{bosons} 
of constant interaction potential, $V(\epsilon,\epsilon')=V$. Then, 
the direct and mutual exclusion statistics parameters
are \cite{submitted.FESinteraction}
\begin{subequations} \label{alpha2}
\begin{equation}
\tilde\alpha_{\tilde\epsilon\tilde\epsilon} = \alpha_{\tilde\epsilon\tilde\epsilon} = V\sigma[\epsilon(\tilde\epsilon)],
\label{direct_alpha}
\end{equation}
and 
\begin{equation}
\tilde\alpha_{\tilde\epsilon\tilde\epsilon'} = 
\theta(\tilde\epsilon-\tilde\epsilon')V\left.\frac{d\{\ln[\sigma(\epsilon)]\}}
{d\epsilon}\right|_{\epsilon(\tilde\epsilon)}\cdot\sigma[\epsilon(\tilde\epsilon)]
\delta\epsilon \equiv \alpha_{\tilde\epsilon\tilde\epsilon'}\delta G(\epsilon) , 
\label{mutual_alpha}
\end{equation}
\end{subequations}
respectively, where $\delta\epsilon$ is the energy interval in which 
the mutual statistics is manifested and 
$\delta G=\sigma[\epsilon(\tilde\epsilon)]\delta\epsilon$ is the 
number of energy levels contained in it. 
Having the exclusion statistics parameters (\ref{alpha2}), we can write 
the integral equation (19) of Ref. \cite{JPhysA.40.F1013.2007.Anghel} 
for the most probable particle population as: 
\begin{equation}
\beta[\mu-\tilde\epsilon(\epsilon)]+\ln\frac{[1+n(\epsilon)]^{1 
-V\sigma(\epsilon)}}{n(\epsilon)} = 
V\int_{\epsilon}^\infty\ln[1+n(\epsilon')]\left[\frac{d\sigma}{d\epsilon'}\right]
\rmd\epsilon' . \label{pop_int}
\end{equation}
Differentiating both sides of (\ref{pop_int}) with respect to $\epsilon$, 
we obtain the differential equation, 
\begin{equation}
\frac{dn}{d\epsilon}\cdot\frac{1+Vn(\epsilon)\sigma(\epsilon)}
{n(\epsilon)[1+n(\epsilon)]} = -\beta\frac{d\tilde\epsilon}{d\epsilon}.
\label{eq_dif_1}
\end{equation}
If we use (\ref{epstilgenint}) into (\ref{eq_dif_1}), the differential 
equation reduces to 
\begin{equation}
\beta^{-1}\frac{dn(\epsilon)}{d\epsilon} = -n(\epsilon)[1+n(\epsilon)] , 
\end{equation}
which, provided that $n(\epsilon)$ is positive and converges to zero 
at $\epsilon\to\infty$, admits only the solution 
\begin{subequations} \label{pop_eqs}
\begin{equation}
n(\epsilon) = \{\exp[\beta(\epsilon-\mu')]-1\}^{-1} . \label{Bose_pop}
\end{equation}
This is, of course, the Bose population in the variable $\epsilon$, 
and relation (\ref{Bose_pop}) is true for any $\sigma(\epsilon)$ and 
constant $V$. To determine $\mu'$, we plug 
(\ref{Bose_pop}) back into (\ref{pop_int}) and obtain
\begin{equation}
\mu' = \mu-VN , \label{eq_mup}
\end{equation}
\end{subequations}
where $N=\int_0^\infty\sigma(\epsilon)n(\epsilon)\rmd\epsilon$ is the total 
number of particles in the system. From $n(\epsilon)$, the population 
$n(\tilde\epsilon)$ follows by a simple change of variable, given by 
equation (\ref{epstilgen}). 

Equations (\ref{pop_eqs}) may seem surprising, but let's have now another 
perspective on the problem. Since 
$V$ is a constant, the total energy of the system reduces to 
\begin{equation}
E = \sum_i \epsilon_i n_i +\frac{VN^2}{2} , \label{E_Vconst}
\end{equation}
which is the energy of the gas in the mean-field approximation. This case is 
easy to treat also from the perspective of the free particle energies, 
$\epsilon$. If we turn back to the division of the energy axis 
$\epsilon$ into the intervals $[\epsilon_i,\epsilon_{i+1})$, $i=0,1,\ldots$, of 
$N_i=\int_{\epsilon_i}^{\epsilon_{i+1}}n(\epsilon)\sigma(\epsilon)\rmd\epsilon$ 
particles and $G_i=\int_{\epsilon_i}^{\epsilon_{i+1}}\sigma(\epsilon)\rmd\epsilon$ 
states, we write the partition function and we maximize it with respect 
to the populations $n(\epsilon)$, then we get 
\begin{subequations}\label{MFA}
\begin{equation}
n[\epsilon'(\epsilon)] = \{\exp[\beta(\epsilon'-\mu)]-1\}^{-1} , 
\label{Bose_popP}
\end{equation}
which is exactly equation (\ref{Bose_pop}), but with a redefinition of the 
quasiparticle energy, 
\begin{equation}
\epsilon'=\epsilon+VN . \label{qpen}
\end{equation}
\end{subequations}
The quasiparticle energies 
(\ref{qpen}) are different from the ones defined in (\ref{epstilgen}). 
Moreover, $\epsilon'_i-\epsilon_i$ is independent of $i$, 
unlike $\tilde\epsilon_i-\epsilon_i$, which is 
$\sum_{j=0}^{i-1} V_{ij}n_j +\frac{1}{2}V_{ii}n_i$ and depends on $i$. 
Another difference is that while the total energy of the system can be written 
as $E=\sum_i\tilde\epsilon_i n_i$, the summation $\sum_i\epsilon'_in_i$ 
gives $E+VN^2/2$. 

So, the definition of quasiparticle energies (\ref{epstilgen}) leads to 
the manifestation of FES along the quasiparticle energy axis, 
$\tilde\epsilon$. If $V$ is constant, the FES particle distribution, 
$n(\tilde\epsilon)$, corresponds to the Bose distribution in the 
free-particle energies, $n(\epsilon)$ (equations \ref{pop_eqs}). 
If $\sigma(\epsilon)$ is also constant, then the mutual exclusion statistics 
parameters, $\alpha_{\tilde\epsilon\tilde\epsilon'}$ (equation \ref{mutual_alpha}) 
vanishes and the direct exclusion statistics parameters become independent 
of $\tilde\epsilon$: $\alpha_{\tilde\epsilon\tilde\epsilon}=V\sigma$; 
if $V\sigma=1$, then the interacting Bose gas may be interpreted as a 
Fermi gas. From here, the thermodynamic equivalence of systems of the same, 
constant density of states and any exclusion statistics \cite{PhysRevE.55.1518.1997.Lee,PhysRevA.63.035601.2001.Crescimanno,JPA35.7255.2002.Anghel} follows directly. 

\section{CONCLUSIONS}

This paper is a continuation of Ref. \cite{submitted.FESinteraction}, where 
we showed that the fractional exclusion statistics (FES) is manifested 
in general in systems of interacting particles and the mutual exclusion 
statistics parameters are proportional to the dimension of the Hilbert space 
on which they manifest. Here we calculated the density of quasiparticle 
states in general and we analyzed in particular a system in which the 
inter-particle interaction potential does not depend on the particles 
quantum numbers ($V_{ij}\equiv V$). Using the formalism presented in 
Refs. \cite{JPhysA.40.F1013.2007.Anghel,submitted.FESinteraction}, we 
showed that in a gas of interacting bosons, the resulting FES quasiparticle 
population, obtained by 
the maximization of the partition function, is actually the original 
Bose distribution written as a function of the quasiparticle energy, 
$\tilde\epsilon$, instead of the free-particle energy, $\epsilon$. 
Therefore the FES reduces to a change of variable, from $\epsilon$--the 
free-particle energy--to $\tilde\epsilon$--the quasiparticle energy--in 
the population of the single particle energy levels, $n$. Form 
this result, the thermodynamic equivalence of systems of the same, constant, 
density of states but any exclusion statistics \cite{PhysRevE.55.1518.1997.Lee,PhysRevA.63.035601.2001.Crescimanno,JPA35.7255.2002.Anghel} follows immediately. 

\subsubsection*{Acknowledgments}

The work was partially supported by the NATO grant, EAP.RIG 982080. 

\section{REFERENCES}

\end{document}